\documentclass[12pt,letterpaper]{article}
\usepackage[letterpaper,left=3cm,right=3cm,top=3cm,bottom=3cm]{geometry}
\usepackage{hyperref}
\usepackage[utf8]{inputenc}
\usepackage{amsmath}
\usepackage{amsfonts}
\usepackage{amssymb}
\usepackage{dsfont}
\usepackage{empheq}
\usepackage{array}
\usepackage{color}
\usepackage{authblk}
\title{The variational approach to infrared divergencies: Applications to black hole action integrals and holographic Wilson loops}
\author{M. Arroyo, D. Arteaga, M. Bañados and A. Faraggi }
\affil{ {\it Facultad de F\'isica, Pontificia Universidad Cat\'olica de Chile,  Santiago, Chile}}
\begin{document}
\maketitle
\begin{abstract}
Renormalizing the on-shell action {\it variation} offers two important advantages over renormalizing the action itself. First, the variation of the action has a universal form in which the bulk contribution vanishes identically. As a result, the renormalization problem is automatically localized at the boundary;  no bulk subtractions are required. Second,  the variation of the action admits a natural interpretation as a 1-form on functional space. The distinction between exact and non-exact 1-forms provides a clear renormalization prescription.

We apply these ideas to the computation of on-shell black hole actions. After presenting the general algorithm, we discuss some examples, including  Kerr–Newman, Kerr–Newman–AdS, Taub–NUT, Taub–Bolt, and STU black holes. We also apply the same techniques to the computation of holographic Wilson loops. All results agree with standard results, in particular with holographic renormalization. 

\end{abstract}
\tableofcontents
\section{Introduction}
In field theory, the evaluation of actions on solutions, the on-shell action,  is a subtle process. For black hole thermodynamics this was noted already in the classic Gibbons-Hawking paper \cite{Gibbons:1976ue}. Two classic examples are: (i) plug the Schwarzschild solution into $\int \sqrt{g}R$. Since the Schwarzschild metric has $R=0$ the action is zero. (ii) plug Schwarzschild-AdS into $\int \sqrt{g}(R+ 2\Lambda)$. Since Schwarzschild-AdS has  $R=\Lambda$, replacing in the action on gets $ \Lambda \int \sqrt{g} = \infty$!  The two most basic examples show that there is something that needs to be understood. The solution to these `paradoxes' is to properly set up the variational problem adding the correct boundary terms to the action, as well as subtracting the IR divergencies. Computing the on-shell value of an action is the art of identifying the correct boundary term. At the risk of being repetitive, $\int \sqrt{g}R$  is not the right action for the Schwarzschild metric, and $\int \sqrt{g}(R+ 2\Lambda)$ is not the right action for the Schwarzschild-AdS metric. 

In this paper we discuss a variational approach to compute on-shell actions. This method can be applied to any problem, although in this paper we concentrate on black hole physics. The method is algorithmic, following a series of steps one obtains the right action for any problem.  The key step is to renormalize the {\it variation} of the action  rather than the action itself. One crucial aspect of looking at the action variation is that all infrared divergences appear only through exact variations \cite{Andrade:2006pg} (we provide a different proof of this result below), thus they can be removed without even computing them. In particular, no background subtraction is necessary. The finite on-shell action is reconstructed by integrating the renormalized variation. The method is covariant and applicable to a wide class of systems, independent of the form of the Lagrangian, including, for example, theories with Chern--Simons interactions. Of course, our results coincide with classic results, in particular holographic renormalization. Examples of the variational approach has been discussed before (see the end of this section). In this paper we provide a systematic algorithm highlighting the generality of the approach. 

The variational method differs substantially from most treatments discussed in the literature. The solution proposed in \cite{Gibbons:1976ue} introduced a boundary term, the Gibbons-Hawking term, motivated by a careful formulation of the variational problem. Although divergent, this term provided the right answer after subtracting the background action (adjusting the proper time period \cite{Witten:1998zw}). Later, studying time evolution in the Euclidean manifold \cite{Hawking:1980gf}, Hawking himself pointed out that the Schwarzschild black hole action was $\frac{1}{2}\beta M$ and not $\beta M$ due to a boundary term at the horizon. The term $\frac{1}{4}A$ as a boundary contribution of the horizon first appeared in \cite{Brown:1990fk,Brown:1992bq}. Generalizations to other dimensions and the general Lovelock action  were studied in \cite{Jacobson:1993xs},\cite{Banados:1992wn},\cite{Banados:1993qp}.

A big step towards the understanding of black hole entropy and the corresponding first law was taken in \cite{Wald:1993nt,Iyer:1994ys}, where a large class of Lagrangians was studied. The starting point in these works is a covariant expression for the (variation of the) charges carried by the solutions ($C$ is a Cauchy surface),
\begin{empheq}{alignat=7}\label{CC}
    \delta q_n&=\int_{C}dC_\mu\,J_n^\mu(\phi,\delta \phi)\,.  
\end{empheq}
Here, $J_n^\mu$ are conserved currents depending on the fields and their variations. Assuming the existence of a stationary black hole with a bifurcate horizon, the first law emerges and a formula for the entropy is identified \cite{Wald:1993nt,Iyer:1994ys}. This is done in great generality with a minimum set of conditions.

The Euclidean Hamiltonian formalism \cite{Brown:1989fa,Brown:1992bq,Brown:1992br,Banados:1992wn} also provides a general proof of the first law of black hole thermodynamics. The Euclidean Hamiltonian action for a diffeormorphism-invariant theory has the general form
\begin{eqnarray}
I_H = \int dt \int d^{d-1}x \Big( \pi_I \dot \phi^I - \lambda^\alpha \psi_\alpha \Big)\ + \ \beta (M + \mu^n q_n)\Big|^{\infty}  - \ B_+\Big|_{r_+}  \label{IH},
\end{eqnarray} 
where $\pi_I,\phi^J$ are the canonical pairs of the theory (including the metric and its momentum), $\psi_\alpha$ are constraints (first and/or second class) and $\lambda^\alpha$ their Lagrange multipliers. The total mass $M$ and the charges $q_n$ appear in the action as boundary terms at infinity. They are accompanied by the corresponding chemical potentials $\beta$ and $\mu^n$, which are needed to have a regular solution in the bulk. In contrast, $B_+$ is a boundary term at the horizon $r=r_+$ and depends on the local geometry at that point. It is added so that the variation of the action receives no contribution from the horizon and can be written as
\begin{eqnarray}
\delta I_H =\left[\textrm{eom}\right] + M \delta \beta + q_n \delta (\beta \mu^n)\,, \label{dIH}
\end{eqnarray}
where [eom] is the bulk piece proportional to the equations of motion. Going on-shell, equating the variation of \eqref{IH} with \eqref{dIH} for a stationary black hole gives the first law
\begin{empheq}{alignat=7}
	\delta B_+&=\beta(\delta M+\mu^n\delta q_n)\,.
\end{empheq}
The boundary term $B_+$ is then identified with entropy. We must mention that, although this analysis is valid in general, it is not always easy to write the action in Hamiltonian form (e.g. higher curvature theories).

An entirely new line of developments emerged with the discovery of the AdS/CFT correspondence \cite{Maldacena:1997re,Witten:1998qj,Gubser:1998bc}. The duality between a gravitational theory in the bulk and a boundary field theory motivated a different, background independent way of subtracting infinities from on-shell actions \cite{Balasubramanian:1999re}. A full analysis of black hole thermodynamics for Einstein-Hilbert gravity coupled to matter fields was developed in \cite{Papadimitriou:2005ii}, along the lines of Holographic Renormalization \cite{deHaro:2000vlm}. The counterterm method is particularly attractive because the very same counterterms required to compute CFT anomalies and correlators also yield the correct action and entropy for black holes. This approach was successfully applied, among other examples, to the calculation of the action for the Taub-NUT solution \cite{Chamblin:1998pz,Emparan:1999pm,Astefanesei:2004ji}, where previous methods had failed due to the nontrivial topology.

~

The variational method is not exclusive to black holes and can be implemented in a variety of problems. It first appeared in \cite{Banados:2004zt} in the context of holographic anomalies in Chern-Simons gravities. In that work, the four-dimensional gravity holographic stress tensor was also computed, with results fully consistent with \cite{Balasubramanian:1999re,deHaro:2000vlm}. A more general analysis of holographic correlators was later presented in \cite{Andrade:2006pg}. The main point emphasized in those references is that, within the variational approach, correlators can be computed directly from the asymptotic solutions without the need to invert the series. In the present paper, we apply the same ideas to the computation of black hole actions. To further illustrate the generality of the method, we also include two examples involving holographic Wilson loops.

Additional examples were developed in \cite{Banados:2012ue} and \cite{Bahamonde:2025qtc} (we do not repeat them here). In \cite{Banados:2012ue} the action and entropy for 3d higher spins black holes  \cite{Gutperle:2011kf}  was computed using a variational approach. This is relevant because, for Chern-Simons interactions, Wald's entropy formula, for example, does not apply.  In  \cite{Bahamonde:2025qtc} the free energy and entropy for a black hole with Gauss-Bonnet interactions and non-metricity was worked out.

The paper is organized as follows. In section \ref{sec: method} we present the variational method. Section \ref{sec: examples} is devoted to a series of examples illustrating the scope and simplicity of the method. We begin with the Schwarzschild black hole, where the issue is purely variational and no infrared divergences arise. We then study the Schwarzschild-AdS solution, where both infrared divergences and variational subtleties are present. The Taub-NUT-AdS and Taub-Bolt-AdS geometries provide more nontrivial examples due to their fibered topology. Since our method does not rely on bulk subtraction, it applies naturally to these spaces and reproduces the results obtained by holographic renormalization \cite{Emparan:1999pm}. These first examples share the property of carrying a single conserved charge, the total energy $M$, so that the renormalized action satisfies the simple variational relation $\delta I=M\delta\beta$.\footnote{The Hamiltonian action also has this form. The main advantage of our (Lagrangian) approach is that we easily renormalize away all IR divergencies, a problem that can be quite subtle in the Hamiltonian formalism.} Regularity conditions determine $M$ as a function of the inverse temperature $\beta$, allowing the action to be reconstructed directly by integration. We also discuss the JT black hole, which illustrates how the variational principle reacts in the presence of trivial parameters in the solution. The analysis is then extended to systems with multiple charges, including Kerr-Newman, Kerr-Newman-AdS, and STU black holes. In these cases, the integrability of the action imposes nontrivial consistency conditions on the charges, making the consistent elimination of conical and other singularities essential. Finally, we conclude with examples of a different nature, namely the computation of holographic Wilson loops, where the same variational ideas provide a direct and transparent treatment of both finiteness and boundary contributions.
\section{The method}\label{sec: method}
In this section, we present the algorithm to compute the renormalized free energy and entropy of black hole solutions. We restrict our attention to Euclidean manifolds with black hole topologies having a single boundary at infinity.

The method has three stages. First, study the variation of the action evaluated on the space of asymptotic solutions and isolate its non-exact part. Second, observe that all infrared divergences appear as exact variations and can therefore be removed without affecting the finite non-exact contribution. Finally, regularity conditions on the horizon render the remaining variation integrable, allowing the finite action to be reconstructed directly. 

We now discuss each step in detail. It should be clear by the presentation that several of the results presented here are properties of field theories, not necessarily black holes.  
\subsection{Action variation as a 1-form}
Let $\phi^a(x)$ be a collection of fields, including the metric, that satisfy the equations of motion coming from a covariant action
\begin{empheq}{alignat=7}
    I_0[\phi]&=\int_M d^dx \, \mathcal{L}(\phi)\,.
\end{empheq}
For example, it could be a black hole solution with matter. In order to compute the on-shell action, one normally seeks a boundary term $B[\phi]$ such that
\begin{empheq}{align=7}\label{I0}
    I[\phi]&=I_0[\phi]+B[\phi]
\end{empheq}
is finite and has well-defined variations. This problem can be quite cumbersome and the details vary significantly from case to case. So, instead of an upfront calculation of $I[\phi]$, we will look at the variation,
\begin{empheq}{align=7}\label{di0}
    \delta I[\phi]&=\delta I_0[\phi]+\delta B[\phi]\,,
\end{empheq}
and construct a boundary term $\delta B[\phi]$ such that $ \delta I[\phi]$ is finite and depends on the correct variables. Once the renormalized form of $\delta I[\phi]$ is known, the actual value of the action can be computed via integration. This last step is subtle, we discuss the details below. 

This approach has several advantages. First, since we are dealing with $\delta I[\phi]$, the variational problem is under control all throughout the calculation. Second, unlike the action itself, the variation of the act ion has a universal form. For any action $I[\phi]$ its variation reads,
\begin{empheq}{alignat=7}\label{dI0}
    \delta I[\phi]&=\int_M d^dx\,E_a\delta\phi^a+\int_{\partial M}d\Sigma_{\mu}\,\theta^{\mu}_a(\phi)\delta\phi^a+\delta B[\phi]\,,
\end{empheq}
where $E_a$ is proportional to the equations of motion and $\theta^{\mu}_a$ is generated in the process of integrating by parts.
This will allow us to make statements valid for any theory that descends from a local Lagrangian. For example, the bulk piece in $\delta I[\phi]$ vanishes on-shell, so all divergences are located at the regulated boundary $\partial M$ and do not come from bulk integrals. This means that no background subtraction is necessary to renormalize the variation of the action, only counterterms defined locally at the boundary (as in holographic renormalization). Finally, variations can be understood as differential forms in functional space, and it is this framework that will enable us to prove that divergences can be easily removed, without computing them, yet keeping all information about the finite part.

Before dealing with divergencies, let us explore some of the consequences of looking at \eqref{dI0} as a 1-form in the space of fields \cite{Crnkovic:1986ex}. This interpretation only makes sense on-shell, such that the variations $\delta\phi^a$ are tangent to the surface $E_a=0$ and satisfy the linearized equations of motion. For simplicity, we will work with solutions characterized by a finite number of constant parameters (charges and chemical potentials), so the configuration space is actually finite-dimensional and the operation $\delta$ becomes the usual exterior derivative $d$.

By definition, the left-hand side of \eqref{dI0} is an exact form, so the right-hand side must be as well. For $E_a=0$ we are left with the 1-form $\theta^{\mu}=\theta^{\mu}_{a}(\phi)\delta \phi^a$, which, before integrating, is not expected to be exact. Moreover, the 2-form $\omega^\mu=\delta\theta^\mu$ is conserved and its integral defines the so-called symplectic structure. On a manifold with two boundaries the sympletic structure is equal on both surfaces. But Euclidean black holes have a single boundary (the other boundary can be shrunk to the horizon). One concludes that $w^{\mu}$ must vanish and there is no globally defined canonical structure.\footnote{In the Hamiltonian formalism, one normally removes a little disk around the horizon thus changing the topology. But, strictly speaking, the Euclidean black hole manifold should be understood as having a single outer boundary \cite{Gibbons:1976ue}.} 

So, why is $\theta^\mu$ exact for black hole solutions? The key observation is that the equations of motion $E_a=0$ are valid \emph{everywhere} only if the fields satisfy specific regularity conditions. In other words, the bulk term in \eqref{dI0} does not vanish for singular configurations. In turn, these regularity conditions restrict the number of free parameters and, only then, does $\theta^\mu$ become exact. In that case there exists a function $F^{\mu}(\phi)$ such that
\begin{empheq}{alignat=7}
    \delta I[\phi]&=\int_{\partial M}d\Sigma_{\mu}\delta F^{\mu}(\phi)+\delta B[\phi]\,,
\end{empheq}
and the on-shell value of the action is (up to constant terms with no variation)
\begin{empheq}{alignat=7}
    I[\phi]&=\int_{\partial M}d\Sigma_{\mu}F^{\mu}(\phi)+B[\phi]\,.
\end{empheq}
This is the main mechanism that will allow us to evaluate black hole actions. Still, many details needs to be clarified; we have said nothing about the boundary term $B$, nor possible IR divergencies. The variational principle and boundary conditions now come into the scene.
\subsection{Euclidean black hole action and first law}

We now focus specifically on black holes. Our discussion is however general, no particular theory/solution is singled out.

Our starting point is somehow the inverse of equation \eqref{CC}. Instead of writing the charges in terms of the fields and their variations, we will write the fields in terms of the charges (and chemical potentials). For our purposes, a black hole is a set of fields
\begin{empheq}{alignat=7}\label{osf}
    \phi^a(x,q_n,\mu_m)\,,&
    &\qquad
    n,m&=1,2,\ldots,N\,,
\end{empheq}
characterized by $N$ charges $q_n$, including the total energy $M$, and $N$ chemical potentials $\mu_n$, one of which is the period $\beta$. We shall also need the variations
\begin{empheq}{alignat=7}\label{dosf}
    \delta\phi^a(x,q,\mu)&=\frac{\partial\phi^a(x,q,\mu)}{\partial q_n}\delta q_n+\frac{\partial\phi^a(x,q,\mu)}{\partial\mu_n}\delta\mu_n
\end{empheq}
which are tangent to the space of solutions \eqref{osf} and solve the linearized equations. For example, the Euclidean Reissner-Nordstr\"om black hole is understood as the pair of fields
\begin{empheq}{alignat=7}\label{A0}
    ds^2&=\left(1-\frac{2M}{r}+\frac{Q^2}{r^2}\right)\beta^2dt^2+\frac{dr^2}{{\displaystyle1-\frac{2M}{r}+\frac{Q^2}{r^2}}}+r^2d\Omega^2\,,
    \cr
    A&=\left(\mu+\frac{Q}{r}\right)\beta\,dt\,,
\end{empheq}
characterized by the 4 parameters $\{M,Q,\beta,\mu\}$. In this paper we work exclusively in Euclidean signature. Parameters like electric charge, angular and their chemical potentials will need factors of $i$ when going back to Minkowski. The Kerr-Newman solution will have six parameters $\{M,J,Q,\beta,\Omega,\mu\}$, and so on. 

Notice that the chemical potentials appear at leading order, whereas the charges are subleading. This is a general feature. In the language of AdS/CFT, they correspond to, respectively, non-normalizable (sources) and normalizable modes (vevs).

Chemical potentials are needed to ensure regularity of the solution. One often hides them via coordinate redefinitions/gauge choices. For example, $\beta$ can be encoded in the range of the Euclidean time coordinate $t$. The problem is that the integration limit in the action then depends on a quantity that has non-vanishing variation. This can lead to confusion. In order to avoid this, we assume that the coordinates take values on fixed ranges and that all physical parameters are included explicitly in the solution.

In this paper, we shall not elaborate on two important aspects of black holes that have been discussed at length in the literature. First, for the definition of charges important references are \cite{ReggeTeitelboim},\cite{Iyer:1994ys},\cite{Barnich:2001jy},\cite{Barnich:2004uw}. See \cite{Gibbons:2004ai} for a detailed analysis of different currents, their charges (for Kerr black holes) and the validity of the first law. With a pragmatic view, the charges (and  chemical potentials) arise as integration constants in the on-shell fields. Of course there is a huge ambiguity in the parametrization of these constants and for that reason associating them to conserved currents, as in (\ref{CC}), is mandatory, but here we assume that this step has already been done. Having said that, it is worth pointing out that the action does not respond to variations of trivial constants: Our method is sensitive to the invariant meaning of the parameters in (\ref{osf}).  We will see this effect in the JT black hole example. Ideally, one would like to have expressions for all $q_n$ in terms of the fields. But, our goal is to display general properties of Lagrangians valid in all situations: higher derivatives, Chern-Simons interactions, matter fields, gauge fields, etc. Is this broad picture it seem rather difficult to display explicit formulas for the charges.   Our main goal is then, for a given solution (\ref{osf}), we would like to compute its renormalized on-shell action and entropy in a quick way. A second aspect that we shall not discuss is the global definition of horizons. We assume that the field (\ref{osf}) has a horizon and its Euclidianization has a black hole topology, in particular, asymptotically, it looks like $\Re\times K $ with $K$ a compact manifold with no boundary.

~

Consider now the variation of the action \eqref{dI0} evaluated on the black hole solution \eqref{osf} and \eqref{dosf}, and focus on the boundary integral involving $\theta^{\mu}$. We introduce a large but finite regulator $r$ where the boundary $\partial M$ is defined. At this point we assume that the equations of motion have only been solved asymptotically for $r\to\infty$, to sufficiently high orders, without imposing any regularity conditions in the interior. In particular, charges and chemical potentials are independent. Our main result can  be stated as the following: the boundary term ($\theta^\mu$-contribution) in \eqref{dI0} has the universal form,
\begin{empheq}{alignat=7}\label{exp}
    \int_{\partial M}d\Sigma_\mu\theta^\mu_a(\phi)\delta\phi^a&=q_n\delta\mu_n+\delta\left(\Theta_0(q,\mu)+\Theta_{\infty}(r,\mu)\right)\,.
\end{empheq}
This expression is justified as follows:
\begin{enumerate}
    \item To reach (\ref{exp}) we assume that the integrals over the coordinates on $\partial M$ have been performed, then the right hand side depends only on the regulator $r$, the charges $q_n$ and their variations $\delta q_n$, as well as the chemical potentials $\mu_n$ and their variations $\delta\mu_n$. In practical terms, we have pullback-ed the symplectic potential $\theta^\mu(\phi,\delta\phi)$ to the subspace of solutions \eqref{osf} characterized by $2N$ real constants.
    
    \item The terms $q_n\delta \mu_n$ and $\Theta_0(q,\mu)$ are both $r$-independent (``zero modes") and hence finite. The function $\Theta_0(q,\mu)$ is chosen such that the non-exact part of this one-form is precisely $q_n\delta \mu_n$, as appropriate for a canonical ensemble. The existence of such a function is guaranteed by Darboux's theorem. Of course, the method is applicable to other ensembles. Once the ensemble has been chosen, $\Theta_0(q,\mu^n)$ is uniquely defined up to a constant with vanishing variation. The appearance of this finite term, which is related here to Darboux's theorem, is fully consistent with the finite terms encountered in standard holographic renormalization \cite{deHaro:2000vlm,Papadimitriou:2005ii}.
    
    \item The most important feature, proved in section \ref{AEDE}, is that {\it all} IR divergencies are {\it exact variations}, that is, they always appear under the sign $\delta$. We collect them in $\delta\Theta_\infty(r,\mu)$.  Thus, \eqref{exp} is naturally (and universally) split into a {\it finite} non-exact form $q_n\delta \mu_n$ plus a divergent exact form. We are also asserting that the charges $q_n$ do not appear in the divergent contribution $\Theta_\infty(r,\mu)$. In the context of AdS/CFT, this corresponds to the statement that the counterterms are local functions of the sources. Although we do not have a general proof of this assertion, it holds in all the examples discussed below. This is closely related to the distinction between normalizable and non-normalizable modes, since one does not expect the former to contribute to infrared divergences.
\end{enumerate}

Given the above decomposition, our renormalization prescription for the variation of the action is now clear: we choose $B[\phi]$ in \eqref{dI0} such that all \underline{exact} forms, finite and divergent, are subtracted, that is,
\begin{empheq}{alignat=7}\label{cb}
    B&=-\left(\Theta_0(q,\mu)+\Theta_\infty(r,\mu)\right)\,.
\end{empheq}
The usual ambiguity associated with adding a finite contribution to the divergent piece is now irrelevant. Such redefinitions of $\Theta_{\infty}$ and $\Theta_0$ do not affect the non-exact term $q_n\delta\mu_n$, from which the value of the action is extracted. With this choice, the variation $\delta I[\phi]$ becomes finite and uniquely defined,
\begin{empheq}{alignat=7}\label{exp2}
    \delta I[\phi]&=\int d^dx\,E_a\delta\phi^a+q_n\delta\mu_n\,.
\end{empheq}
The calculation of $I[\phi]$ is nostraightforward. 

In order to integrate the variation we impose the equations of motion everywhere, not just asymptotically.  Regularity of the solution at the horizon implies exactly $N$ algebraic relations between charges and chemical potentials, fixing $q_n=q_n(\mu)$. We reiterate that if these conditions are not implemented the equations $E_a=0$ are not globally  satisfied. Thus, setting $E_0=0$  has two effects on \eqref{exp2}, the bulk term vanishes and the boundary term becomes exact, that is,
\begin{empheq}{alignat=7}\label{exp3}
    \delta I[\phi]&=q_n(\mu)\delta\mu_n\,,
\end{empheq}
where the functions $q_n(\mu)$ satisfy the integrability conditions
\begin{empheq}{alignat=7}\label{consis0}
    \frac{\partial q_n(\mu)}{\partial\mu_m}&=\frac{\partial q_m(\mu)}{\partial\mu_n}\,.
\end{empheq}
The integral over $\mu_n$ is path independent and the finite renormalized action (in the canonical ensemble) becomes
\begin{empheq}{alignat=7}\label{exp4}
    I(\mu)&=\int q_n(\mu)d\mu_n\,.
\end{empheq}
It is not at all obvious that (\ref{consis0}) is equivalent to imposing regularity at the horizon, but it follows from the fact that \eqref{exp3} is a total variation, leading to the existence of the action $I[\phi]$. See \cite{Gutperle:2011kf} for an example in which the action in computed using only integrability conditions.

Notice that in order to reach the final renormalized action $I[\mu]$ in \eqref{exp3}, we do not need to compute the IR divergencies explicitly. We only need to know that they are exact variations and thus can be subtracted away. For any theory admitting black hole solutions, the action appropriate for the variational problem with fixed potentials will satisfy \eqref{exp3}. This is valid for any Lagrangian, including Chern-Simons terms, higher spin gravity \cite{Gutperle:2011kf}, etc. 

Starting from Eq. \eqref{exp3}, the calculation of the action and entropy is similar to the Hamiltonian approach \cite{Brown:1989fa,Brown:1992bq,Brown:1992br,Banados:1992wn,Banados:1993qp}. Our point has been a universal treatment of divergencies implying the general validity of (\ref{exp3}) directly from the Lagrangian. In particular, we do not foliate the spacetime nor do we add any boundary terms at the horizon.

To conclude, let us review how the first law follows from \eqref{exp3}. First we need to separate the pair $M,\beta$ from the rest of the charges and chemical potentials. We also redefine the remaining potentials by $\mu_i\to\beta\mu_i$, with $i=1,\ldots,N-1$. The action variation now reads
\begin{empheq}{alignat=7}\label{deltaIos2}
    \delta I[\beta,\mu]&=M\delta\beta+q_i\delta(\beta\mu_i)\,.
\end{empheq}
The entropy function $S[M,q]$ is defined as the Legendre transform
\begin{empheq}{alignat=7}
    S[M,q]&=\beta\left(M+\mu_iq_i\right)-I[\beta,\mu]\,.
\end{empheq}
The first law, 
\begin{empheq}{alignat=7}
    \delta S[M,q]&=\beta\left(\delta M+\mu_i\delta q_i\right)\,,
\end{empheq}
follows from \eqref{deltaIos2}. 

\subsection{Asymptotic equations:  divergencies are exact forms}
\label{AEDE}
We now show that on-shell divergencies in the {\it variation} of the action are always exact forms. This result applies to all classes of asymptotic solutions and is a property of Lagrangian theories, not black holes. The only assumption is that the asymptotic topology is $\Re\times K(r)$, where $\Re$ corresponds to the radial coordinate $r$ (the cutoff) and $K(r)$ is a sequence of compact manifolds with no boundary defined at constant $r$. Many Euclidean black holes fall into this case, for example, for the $4d-$Schwarzschild metric has $K=S_1\times S_2$.\footnote{This is the asymptotic topology. The full topology of Euclidean black holes is $\Re_2\times S_{n}$.} See \cite{Banados:2004zt,Andrade:2006pg} for applications in AdS/CFT. In section (\ref{WilsonSec}) we will apply our results to string worldsheets describing holographic Wilson loops.

For any Lagrangian $\mathcal{L}[\phi]$ its variation is
\begin{empheq}{alignat=7}\label{dL}
    \delta{\cal L}[\phi] = E_{a}\delta \phi^a + \partial_\mu( \theta^{\mu}_{a}(\phi)\delta\phi^a)\,.
\end{empheq}
Now integrate this over the compact space $K(r)$ at some fixed large radius $r$. Since the space has no boundary we obtain ($\partial_r$ is normal to $K(r)$)
\begin{empheq}{alignat=7}\label{cr}
    \int_{K(r)}d^nx\,\delta\cal\mathcal{L}(\phi)&=\int_{K(r)}d^nx\,E_{a}\delta \phi^a+\frac{d}{dr}\Theta(\phi,\delta\phi,r)\,,
\end{empheq}
where we have defined
\begin{empheq}{alignat=7}
    \Theta(\phi,\delta\phi,r)&=\oint_{K(r)}d^nx\,\theta^{r}_a(\phi)\delta\phi^a\,.
\end{empheq}
This is precisely the boundary term that appears in the variation of the action \eqref{dI0}. It depends on the fields, their variations and the regulator $r$. Now, assuming the asymptotic equations have been solved to sufficiently high order, equation \eqref{cr} implies
\begin{empheq}{alignat=7}\label{cr2}
    \delta\left[\oint_{K(r)}d^nx\,\mathcal{L}(\phi)\right]&=\frac{d}{dr}\Theta(\phi,\delta\phi,r)\,.
\end{empheq}
One normally reads \eqref{cr2} from left to right. Here, however, we view it the other way around: the left-hand side contains important information about the right-hand side. Indeed, equation \eqref{cr2} tells us that all terms in $\Theta(\phi,\delta\phi,r)$ that depend on $r$ are exact forms. In particular, any term that diverges as $r\to\infty$ is a total variation. This is a non-trivial statement. The expression $\Theta(\phi,\delta\phi,r)$ depends on all $2N$ parameters $(q_n,\mu_n)$ and is, in fact, not exact. What \eqref{cr2} shows is that the $r$-dependent part of $\Theta(\phi,\delta\phi,r)$ is exact, whereas its $r$-independent part is not.

We also learn from \eqref{cr2} that the counterterm necessary to render the variation of the action finite is invariant: it is the variation of the Lagrangian, integrated over $K$ and $r$. This seems obvious, but the point of this analysis is that we have separated the {\it finite} part. 

A further corollary is that, if a solution has zero Lagrangian, then there will be no divergencies in the variation of the action. For example, all vacuum solutions to Einstein equations (without $\Lambda$) have finite action variations (since $R=0$). The same will be true for the JT black hole.  

Finally, note that \eqref{cr2} provides no information about the $r$-independent part of $\Theta(\phi, \delta\phi,r)$. Indeed, generically, additional finite boundary terms will be necessary to enforce the required boundary conditions. This is in full consistency with finite terms obtained through the holographic renormalization procedure \cite{deHaro:2000vlm}.
\section{Examples}\label{sec: examples}
Having explained the method to compute on-shell actions, we now apply it to some examples. We work in the Euclidean sector assuming all fields and parameters are real. Going back to Minkowski signature would require several complexifications. We start with the simplest cases.

\subsection{Systems with one charge}
For systems with a single charge, the total energy, relation \eqref{exp3} is simply
\begin{empheq}{alignat=7}\label{1c}
    \delta I&=M\delta\beta\,.
\end{empheq}
Regularity at the horizon fixes $M(\beta)$. Once this relation is known, the action $I[\beta]$ can be obtained by integration, and the entropy then follows from $S[M]=\beta M-I[\beta]$. Our claim is that one can start directly from \eqref{1c} and avoid dealing with divergences altogether. Nevertheless, it is worthwhile verifying this explicitly case by case.

The starting point in the following examples is the Einstein-Hilbert action with negative cosmological constant (including the flat limit $l\to\infty$)
\begin{empheq}{alignat=7}\label{I0sch}
    I&=-\frac{1}{16\pi}\int_Md^4x\sqrt{g}\left(R+\frac{6}{l^2}\right)+B\,.
\end{empheq}
Its variation reads
\begin{empheq}{alignat=7}\label{varIEH}
    \delta I&=\frac{1}{16\pi}\int_Md^4x\sqrt{g}\left(G_{\mu\nu}-\frac{3}{l^2}g_{\mu\nu}\right)\delta g^{\mu\nu}
    \cr
    &+\frac{1}{16\pi}\int_{\partial M}d^3x\sqrt{h}n_{\mu}\left(\nabla^{\mu}\delta g^{\nu}_{\phantom{\nu}\nu}-\nabla_{\nu}\delta g^{\mu\nu}\right)+\delta B\,,
\end{empheq}
where $h$ is the determinant of the induced metric on the boundary and $n_{\mu}$ is the unit normal. Notice that we do not introduce the Gibbons-Hawking term.
\subsubsection{Schwarzschild}
The Schwarzschild black hole solution is
\begin{empheq}{alignat=7}\label{sch}
    ds^2&=f(r)\beta^2d\tau^2+\frac{dr^2}{f(r)}+r^2d\Omega^2\,,
    &\qquad
    f(r)&=1-\frac{2M}{r}\,.
\end{empheq}
Inserting this into the variation \eqref{varIEH} (for $l\to\infty$) we find
\begin{empheq}{alignat=7}\label{varI0sch}
    \delta I&=\frac{1}{16\pi}\int d^4x\sqrt{g}\,G_{\mu\nu}\delta g^{\mu\nu}+\frac{1}{2}\left(M\delta\beta-\beta\delta M\right)+\delta B\,,
\end{empheq}
where we have used that the Euclidean time coordinate has period $\tau\sim\tau+1$. Notice that there are no divergences, all the boundary contributions are finite. To put this in canonical form we choose
\begin{empheq}{alignat=7}
    B&=\frac{1}{2}\beta M\,,
\end{empheq}
so that
\begin{empheq}{alignat=7}
    \delta I&=\frac{1}{16\pi}\int d^4x\sqrt{g}\,G_{\mu\nu}\delta g^{\mu\nu}+M\delta\beta\,.
\end{empheq}
This confirms that the integration constant $M$ appearing in \eqref{sch} is the total energy. As discussed above, the bulk contribution proportional to the Einstein tensor vanishes only after imposing Hawking's temperature relation $\beta=8\pi M$. If one instead sets it to zero while treating $M$ and $\beta$ as independent variables, a contradiction arises: the left-hand side is an exact variation, whereas the right-hand side is not. For the regular solution, we can easily integrate the variation to find
\begin{empheq}{alignat=7}
    \delta I&=\frac{\beta}{8\pi}\delta\beta
    &\qquad\Rightarrow\qquad
    I[\beta]&=\frac{\beta^2}{16\pi}\,.
\end{empheq}
The Legendre transform gives the entropy $S=\frac{1}{4}A$.
\subsubsection{Schwarzschild-AdS}
This calculation is almost identical to the Schwarzschild black hole, except that now the variation of the action has a divergent term, namely,
\begin{empheq}{alignat=7}
    \delta I&=\frac{1}{16\pi}\int d^4x\sqrt{g}\left(G_{\mu\nu}-\frac{3}{l^2}g_{\mu\nu}\right)\delta g^{\mu\nu}+\frac{1}{2}\left(M\delta\beta-\beta\delta M\right)+\delta\left(\frac{\beta r^3}{2l^2}\right)+\delta B\,.
\end{empheq}
The key observation is that the divergence is a total variation. Choosing
\begin{empheq}{alignat=7}
    B&=-\frac{\beta r^3}{2l^2}+\frac{1}{2}\beta M\,,
\end{empheq}
the action is finite and satisfies the required variational problem. Indeed,
\begin{empheq}{alignat=7}
    \delta I&=\frac{1}{16\pi}\int d^4x\sqrt{g}\left(G_{\mu\nu}-\frac{3}{l^2}g_{\mu\nu}\right)\delta g^{\mu\nu}+M\delta\beta\,.
\end{empheq}
Now, regularity at the horizon imposes the Gibbons-Hawking temperature condition $\beta=4\pi/f'(r_+)$. The problem is slightly more complicated than before because the horizon $r_+$ is defined by a cubic equation. This, however, is only a technical matter that is easily overcome by writing both $M$ and $\beta$ as functions of $r_+$ using $f(r_+)=0$:
\begin{empheq}{alignat=7}
    M&=\frac{r_+}{2}\left(1+\frac{r_+^2}{l^2}\right)\,,
    &\qquad
    \beta&=\frac{4\pi l^2r_+}{l^2+3r_+^2}\,.
\end{empheq}
The renormalized action then becomes
\begin{empheq}{alignat=7}
    \delta I&=\frac{r_+}{2}\left(1+\frac{r_+^2}{l^2}\right)\delta\left(\frac{4\pi l^2r_+}{l^2+3r_+^2}\right)
    &\qquad\Rightarrow\qquad
    I[\beta]&=\frac{\pi r_+^2\left(l^2-r_+^2\right)}{l^2+3r_+^2}\,,
\end{empheq}
where we have fixed $I[0]=0$. As expected, the entropy is $S[M]=\pi r_+^2$.
\subsubsection{Taub-NUT/Bolt-AdS}
Another class of solutions for which the renormalized action satisfies \eqref{1c} is the Taub-NUT/Bolt metric
\begin{empheq}{alignat=7}
    ds^2&=f(r)\left(\beta dt+2n\cos\theta\,d\phi\right)^2+\frac{dr^2}{f(r)}+\left(r^2-n^2\right)d\Omega^2\,,
    \cr
    f(r)&=\frac{r^2-2Mr+n^2}{r^2-n^2}+\frac{r^4-3n^4-6n^2r^2}{l^2\left(r^2-n^2\right)}\,,
\end{empheq}
where $n$ is the NUT charge. This metric apparently carries two charges, $n$ and $M$, but this is not correct if one demands a regular asymptotic geometry. We take a conservative approach and eliminate the Misner string from the beginning by assuming that
\begin{empheq}{alignat=7}\label{ar}
    \beta&=8\pi n\,.
\end{empheq}
Notice that we have not imposed any regularity condition in the interior yet, this is an asymptotic analysis. The variation of the action reads
\begin{empheq}{alignat=7}
    \delta I&=\frac{1}{16\pi}\int d^4x\sqrt{g}\left(G_{\mu\nu}-\frac{3}{l^2}g_{\mu\nu}\right)\delta g^{\mu\nu}+\frac{1}{2}\left(M\delta\beta-\beta\delta M\right)+\delta\left(\frac{\beta r^3}{2l^2}\right)
    \cr
    &-\frac{rn}{2l^2}\left(n\delta\beta+8\beta\delta n\right)+\delta B\,.
\end{empheq}
Now there are two divergent boundary contributions. It is interesting to notice that the linear divergence is not a total variation if we vary $n$ and $\beta$ independently. Only after imposing \eqref{ar} are the asymptotic equations of motion satisfied identically (there are sources at the strings) and our result ``all divergencies are total variations" applies. To avoid introducing sources we have considered the reduced subspace of solutions where \eqref{ar} holds. Choosing
\begin{empheq}{alignat=7}
    B&=-\frac{\beta r^3}{2l^2}+\frac{3n^2\beta r}{2l^2}+\frac{1}{2}\beta M\,,
\end{empheq}
we land in
\begin{empheq}{alignat=7}
    \delta I&=\frac{1}{16\pi}\int d^4x\sqrt{g}\left(G_{\mu\nu}-\frac{3}{l^2}g_{\mu\nu}\right)\delta g^{\mu\nu}+M\delta\beta\,.
\end{empheq}
\subsubsection*{Taub-NUT}
In this case the function $f(r)$ is required to vanish at the origin $r=n$. This fixes
\begin{empheq}{alignat=7}
    M&=n\left(1-\frac{4n^2}{l^2}\right)\,.
\end{empheq}
The renormalized action is found by simple intergation (assuming $I[0]=0$),
\begin{empheq}{alignat=7}
    \delta I&=8\pi n\left(1-\frac{4n^2}{l^2}\right)\delta n
    &\qquad\Rightarrow\qquad
    I[\beta]&=4\pi n^2\left(1-\frac{2n^2}{l^2}\right)\,.
\end{empheq}
The entropy gives
\begin{empheq}{alignat=7}
    S&=4\pi n^2\left(1-\frac{6n^2}{l^2}\right)\,,
\end{empheq}
which agrees with \cite{Emparan:1999pm,Mann:1999pc}. This is a counterexample of the area law, $S\neq\frac{1}{4}A$, as discussed by \cite{Hawking:1998jf,PhysRevD.59.044033}.
\subsubsection*{Taub-Bolt}
This solution has a horizon at $r=r_+$. From $f(r_+)=0$ we can solve for $M$ as a function of $n$ and $r_+$ and write
\begin{empheq}{alignat=7}\label{MTB}
    M&=\frac{r_+^4+\left(l^2-6n^2\right)r_+^2+n^2\left(l^2-3n^2\right)}{2l^2r_+}\,.
\end{empheq}
Upon using \eqref{ar}, the Gibbons-Hawking condition $\beta=4\pi/f'(r_+)$ relates $n$ and $r_+$, yielding
\begin{empheq}{alignat=7}
    r_+&=\frac{1}{12n}\left(l^2-\sqrt{l^4-48l^2n^2+144n^4}\right)\,.
\end{empheq}
Replacing this in \eqref{MTB} one can integrate the variation to find the renormalized action. The result is
\begin{empheq}{alignat=7}
    I[\beta]&=-\frac{\pi}{216l^2n^2}\left(l^6-\left(l^4-48l^2n^2+144n^4\right)^{\frac{3}{2}}\right)+\frac{\pi l^2}{3}
    \cr
    &=\frac{4\pi n}{l^2}\left(Ml^2+3n^2r_+-r_+^3\right)\,.
\end{empheq}
Here we have fixed the additive constant such that the action has a well-defined flat limit $l\to\infty$. The entropy can be expressed as
\begin{empheq}{alignat=7}
    S[M]&=\frac{4\pi n}{l^2}\left(Ml^2-3n^2r_++r_+^3\right)\,,
\end{empheq}
which, again, is a counterexample of the area law. Our results agree with \cite{Emparan:1999pm,Mann:1999pc}.
\subsubsection{Jackiw-Teitelboim}
We now discuss the thermodynamics of the JT black hole
\begin{empheq}{alignat=7}\label{JT}
    ds^2&=f(r)\beta^2dt^2+\frac{dr^2}{f(r)}\,,
    &\qquad
    f(r)&=\frac{r^2}{l^2}+\frac{2br}{l}-a^2\,,
    \cr
    \Phi&=8\pi c\left(\frac{r}{l}+b\right)
\end{empheq}
which is a solution to the two-dimensional theory
\begin{empheq}{alignat=7}\label{JT0}
    I&=-\frac{1}{16\pi}\int d^2\sqrt{g}\,\Phi\left(R+\frac{2}{l^2}\right)+B\,.
\end{empheq}
This system has 4 integration constants $a,b,c$ and $\beta$. Naively, one may think that there are 2 canonical pairs, but this is not correct. The constant $b$ can be set to zero by the translation $r\to r-bl$. Nevertheless, we will keep it to test how the formalism reacts in the presence of a trivial parameter. On the other hand, as in \cite{Maldacena:2016upp}, we shall treat $c$ as a fixed parameter, similar to a coupling constant.\footnote{The constant $c$ can be eliminated by rescaling $r$. One could vary it in the process, only to discover that the action depends on the invariant combination $\beta/c$.}

The variation of the action is
\begin{empheq}{alignat=7}
    \delta I&=-\frac{1}{16\pi}\int d^2x\sqrt{g}\left(E_{\Phi}\delta\Phi+E^{\mu\nu}\delta g_{\mu\nu}\right)
    \cr
    &+\frac{1}{16\pi}\int_{\partial M}dx\sqrt{h}n_{\mu}\left[\Phi\left(\nabla^{\mu}\delta g^{\nu}_{\phantom{\nu}\nu}-\nabla_{\nu}\delta g^{\mu\nu}\right)+\nabla_{\nu}\Phi\,\delta g^{\mu\nu}-\nabla^{\mu}\Phi\,\delta g^{\nu}_{\phantom{\nu}\nu}\right]+\delta B\,.
\end{empheq}
where
\begin{empheq}{alignat=7}
    E_{\Phi}&=R+2\,,
    &\qquad
    E^{\mu\nu}&=\nabla^{\mu}\nabla^{\nu}\Phi-g^{\mu\nu}\nabla^2\Phi+g^{\mu\nu}\Phi\,,
\end{empheq}
are the equations of motion. For the above solution this becomes
\begin{empheq}{alignat=7}\label{deltaIJT}
    \delta I&=-\frac{1}{16\pi}\int d^2x\sqrt{g}\left(E_{\Phi}\delta\Phi+E^{\mu\nu}\delta g_{\mu\nu}\right)+\frac{c}{l}\left(\left(a^2+b^2\right)\delta\beta+\frac{\beta}{2}\delta\left(a^2+b^2\right)\right)+\delta B\,.
\end{empheq}
The first interesting result is that the boundary term is finite. No counterterms are needed. This is expected because the bulk JT action is identically zero on any solution. A second observation is that the variation depends only on the combination $a^2+b^2$, which is invariant under the translation $r\to r-bl$. Consequently, we will rename
\begin{empheq}{alignat=7}\label{Ab}
    a^2+b^2&=\frac{2Ml}{c}\,.
\end{empheq}
The normalization is adjusted so that $M$ is conjugate to $\beta$ and represents the total energy. Choosing the boundary term as
\begin{empheq}{alignat=7}
    B&=-\beta M
\end{empheq}
we get
\begin{empheq}{alignat=7}\label{deltaIfJT}
    \delta I&=-\frac{1}{16\pi}\int d^2x\sqrt{g}\left(E_{\Phi}\delta\Phi+E^{\mu\nu}\delta g_{\mu\nu}\right)+M\delta\beta\,.
\end{empheq}
The last step is to impose the regularity condition and avoid conical singularities at the horizon. Then the bulk term vanishes and we find
\begin{empheq}{alignat=7}
    \beta&=\frac{2\pi l}{\sqrt{a^2+b^2}}
    &\qquad\Rightarrow\qquad
    \delta I&=\frac{2\pi^2cl}{\beta^2}\delta\beta
    &\qquad
    I[\beta]&=-\frac{2\pi^2cl}{\beta}\,.
\end{empheq}
The entropy is defined via the Legendre transform $S=M\beta-I$ and gives
\begin{empheq}{alignat=7}
    S&=\frac{4\pi^2 cl}{\beta}\,,
\end{empheq}
in full agreement with \cite{Maldacena:2016upp}.

\subsection{Systems with several charges: integrability conditions}

We now review more complicated black holes carrying several charges. In these cases there is additional structure to the problem, namely, the integrability conditions \eqref{consis0}. As we will see, these follow directly from the regularity requirements at the horizon.

\subsubsection{Kerr-Newman}
Our first example with multiple charges is the Euclidean Kerr-Newman black hole.  In Boyer-Lindquist coordinates this field reads,
\begin{eqnarray}
            ds^2  &=& \frac{\Delta}{\rho^2} \left[ \beta dt - a \sin^2{\theta} d \tilde\phi \right]^2 +  \frac{\sin^2{\theta}}{\rho^2} \left[ (r^2 - a^2) d \tilde\phi +   + a \beta dt\right]^2+  \frac{\rho^2}{\Delta} dr^2 + \rho^2 d \theta^2 \nonumber
    \\
     A &=&\left( \mu + \frac{Qr}{\rho^2} \right) \beta dt - \frac{ aQr\sin^2{\theta}}{\rho^2}d\tilde\phi,   \label{Kerr-N}
\end{eqnarray}
where
\begin{eqnarray}
     \tilde \phi &=& \phi - \Omega \beta t, \nonumber\\
    \rho^2 &=& r^2 - a^2 \cos^2{\theta}, \nonumber\\ 
    \Delta &=& r^2 - 2M r - a^2 +Q^2, \nonumber\\
    a &=& J/M. \nonumber
\end{eqnarray}
We shall work in the Euclidean sector assuming all fields and parameters are real. Going back to Minkowski signature will require several complexifications\footnote{Often the notation $Q_E=i Q_M$, and the same for other parameter, is used. Since we work in the Euclidean throughout the paper, we avoid this distinction. When going back to Minkowski several charges and chemical potentials will need to be complexified.}. 

The above field depends on 3 charges $\{ M,J,Q \}$ and three chemical potentials $\{\beta,\Omega, \mu\}$, a total of three canonical pairs. The Euclidean action is the Einstein-Maxwell theory,
\begin{eqnarray}
    I[g, A] = -\frac{1}{16 \pi} \int d^4 x \sqrt{g} \left(R + F^{\mu \nu} F_{\mu \nu} \right) + B,
\end{eqnarray}
where the boundary term $B$ will be adjusted to have a canonical ensemble with $\{\beta,\Omega, \mu\}$ fixed. This is another example where we will see  no divergencies. This can be anticipated by the fact that $R=0$ and $\sqrt{g} F^{\mu\nu}F_{\mu\nu}  \simeq {Q^2 \over r^2} + {\cal O}(1/r^3)$ hence the radial part of $\Theta$ dies away at large $r$. 

The variation of the Einstein-Maxwell action --keeping all boundary terms-- is (here $d^3x \equiv dt d \theta d \phi$)
\begin{eqnarray}
\begin{split}
    \delta I =  \mbox{(eom)} - \frac{1}{16 \pi} \int_{\partial M(r)} d^3 x \, \sqrt{g} \, n_\mu  \Big[ D_\nu \delta g^{\mu \nu}-D^\mu \delta g^\nu_{\ \nu}  +   4 F^{\mu \nu} \delta A_\nu \Big] + \delta B  .\label{dkerr}
\end{split}
\end{eqnarray}
Here (eom) is proportional to the equations of motion of this theory, 
\begin{eqnarray}
 G_{\mu \nu} &=& - 2 F_{\mu \lambda} F_\nu^{\ \lambda} + \frac{1}{2} g_{\mu \nu} F^2 , \nonumber\\ 
 D_\mu F^{\mu \nu}  &=& 0. \nonumber
 \end{eqnarray} 
The bulk piece will eventually be zero, our concern is to evaluate the boundary pieces (at large $r$). To this end we need the perturbation fields $\delta g_{\mu\nu}$ and $\delta A_\mu$  in the $r \to \infty$ regime.  These are computed directly from the field (\ref{Kerr-N}) varying all 6 parameters, see Eq. (\ref{dosf}).  The expressions are a bit long so we skip them.  Replacing in (\ref{dkerr}) the integrals over $t$ and the angles can be done. The result is,
\begin{eqnarray}
    \delta I =  \mbox{(eom)} + \frac{1}{2} \left( M \delta \beta - \beta \delta M\right) + J  \delta (\beta\Omega)  + Q  \delta(\beta \mu) + \delta B
\end{eqnarray}
where (eom) are the equations of motion.  As anticipated, there are no divergencies. Again we choose $B = {1 \over 2}M \beta$ to find the desired canonical result
\begin{eqnarray}
    \delta I =  \mbox{(eom)} +   M \delta \beta  + J  \delta (\beta\Omega)  + Q  \delta(\beta \mu) . \label{KerrdI}
\end{eqnarray}
This is the desired relation for the renormalized action variation, derived here directly from the Eistein-Hilbert action. Once (\ref{KerrdI}) has been established the rest is known but we include it for completeness. 

To compute $I$ from $\delta I$ we would need to insert the regular Euclidean black hole, namely, $M,J,Q$ determined in terms of $\beta,\Omega, \mu$ that ensure a regular horizon. These relations are known, 
\begin{eqnarray}
    \beta = \frac{4 \pi ( r_+^2 - a^2)}{r_+ - r_-}, \qquad \Omega = \frac{a}{r_+^2-a^2}, \qquad \mu = \frac{-Q r_+}{r_+^2 - a^2}
\end{eqnarray}
with $r_\pm = M \pm \sqrt{M^2 + a^2 - Q^2}$. We see that it is easy to write $\beta,\Omega,\mu$ it terms of $M,J,Q$, but not the other way around. What we can do is to go directly to  the entropy via the Legendre transform,
\begin{eqnarray}
S = \beta (M+\Omega J + \mu Q) - I
\end{eqnarray} 
satisfying, 
\begin{eqnarray}
\delta S = \beta (\delta M + \Omega \delta J +  \mu \delta Q)
\end{eqnarray} 
The  integrability conditions hold, 
\begin{eqnarray}
 \frac{\partial \beta}{\partial J} =  \frac{\partial{(\beta \Omega)}}{\partial M}, \qquad \frac{\partial \beta}{ \partial Q} = \frac{\partial (\beta \mu)}{\partial M}, \qquad {\partial( \beta \mu )\over \partial J} = {\partial (\beta \Omega) \over \partial Q}
\end{eqnarray}
and the entropy $S(M,J,Q)$ is a function of state,
\begin{eqnarray}
    S &=& 2 \pi \left( M^2 - \frac{1}{2} Q^2 + \sqrt{M^4 - J^2  - Q^2 M^2} \right) \\
    & = &  \pi (r_+^2 - a^2)
\end{eqnarray}
equal to $\frac{A}{4} $, as expected. 

\subsubsection{Kerr-Newman-AdS}

Kerr-Newman-AdS black hole is similar to Kerr-Newman except for the presence of IR divergencies in this case. To exemplify the general result that all divergencies are total variations we treat this example as well. 

In Boyer-Lindquist coordinates the fields read,
\begin{eqnarray}
    & ds^2 =  \frac{\Delta_r}{\rho^2} \left[  \beta dt - \frac{ a \sin^2 \theta}{\Xi } d \tilde{\phi} \right]^2 + \frac{\rho^2}{\Delta_r} dr^2 + \frac{\rho^2}{\Delta_\theta} d \theta^2  + \frac{\Delta_\theta \sin^2 \theta}{\rho^2} \left[ a \beta dt +\frac{r^2 - a^2}{\Xi} d \tilde{\phi}  \right]^2, \nonumber \\
    & A =  \left(\mu + \frac{q r}{\rho^2} \right)  \beta dt-  \left( \frac{  q r a \sin^2 \theta }{\rho^2 \Xi} \right) d \tilde{\phi} \label{KNAdS}
\end{eqnarray}
where 
\begin{eqnarray}
    & \tilde{\phi} = \phi - \Omega_H \beta t, \qquad 
    \rho^2 = r^2 - a^2 \cos^2 \theta, \qquad \Xi = 1 + \frac{a^2}{l^2}, \nonumber \\
     &\Delta_r = (r^2 - a^2) \left( 1 + \frac{r^2}{l^2} \right) - 2mr + q^2, \qquad  
     \Delta_\theta = 1 + \frac{a^2}{l^2} \cos^2 \theta. \nonumber 
\end{eqnarray}
Same as for the case without cosmological constant, these fields depend on 3 charges $\{M,J,Q\}$ and three chemical potentials $\{ \beta ,\Omega, \mu\}$. However, in the fields, the quantities that appear are $m,q,a,\Omega_H$, the actual quantities that enter the thermodynamics are defined by
\begin{eqnarray}
     M =  \frac{m}{\Xi^2}, \qquad J = \frac{am}{\Xi^2}, \qquad Q = \frac{q}{\Xi}, \qquad \Omega = \Omega_H +\frac{a}{l^2}.
\end{eqnarray}
the mass $M$ and the angular momentum $J$ can be defined through the Komar integrals, using the normalized killing vectors $\partial_t/\Xi $ and $\partial_\phi$ and taking the AdS space as reference background. The charge $Q$ is obtained by computing the flux of the electromagnetic field tensor at infinity, and the angular velocity $\Omega$ is defined such that it vanishes at infinity.

The Euclidean action is the Einstein-Maxwell theory with cosmological constant
\begin{eqnarray}
    I[g,A] = - \frac{1}{16 \pi} \int d^4 x \sqrt{g} \left( R - \frac{6}{l^2} + F^{\mu \nu} F_{\mu \nu} \right) + B
\end{eqnarray}
 Similarly to the Schwarzschild-AdS case, the computation of the action is almost identical to the case without cosmological constant. As expected, the variation of the action has a divergent term due to $R = -12/l^2 $, on the other hand, the term proportional to $\sqrt{g}F^{\mu \nu} F_{\mu \nu}$ remains of order $\mathcal{O}(r^{-2})$ so it will not contribute to the divergences. The variation of the action is equivalent to \eqref{dkerr}, and the perturbations $\delta g_{\mu \nu}$ and $\delta A_\mu$ are obtained from \eqref{KNAdS} by varying all of the parameters. Replacing them into the variation of the action and integrating over $t$ and the angles the result is,
\begin{eqnarray}
      \delta I = \mbox{(eom)} 
     + \frac{1}{2} \left( M \delta \beta - \beta \delta M\right) + J \delta( \beta \Omega) +Q \delta(\beta \mu) \nonumber\\
      + \delta \left( \frac{\beta r^3}{2 \Xi l^2 } -\frac{J^2 \beta r}{2M^2 \Xi l^2} - \frac{J^2\beta}{2M l^2}\right) + \delta B
\end{eqnarray}
where $\mbox{(eom)}$ is proportional to the equations of motion, which in this case are given by
\begin{eqnarray}
    G_{\mu \nu} - \frac{3}{l^2} g_{\mu \nu} &=& - 2 F_{\mu \lambda} F_\nu^{\ \lambda} + \frac{1}{2} g_{\mu \nu} F^2 \\
    D_\mu F^{\mu \nu} &=& 0
\end{eqnarray}
The divergence is a total variation, and we choose $B = -\frac{\beta r^3 }{2 \Xi l^2} + \frac{J^2 \beta r}{2M^2 \Xi l^2} + \frac{J^2 \beta}{2M l^2} + \frac{1}{2} \beta M$ to find the desired canonical result
\begin{eqnarray}
    \delta I = \mbox{(eom)} + M \delta \beta + J \delta(\beta \Omega) + Q \delta (\beta \mu)
\end{eqnarray}

To compute the value of the action we replace the regular Euclidean black hole, which is given by the regularity conditions
\begin{eqnarray}
    \beta = \frac{4\pi ( r_+^2 - a^2)}{r_+ \left( 1 - \frac{a^2}{l^2} + 3 \frac{r_+^2}{l^2} - \frac{q^2-a^2}{r_+^2} \right)}, \quad \Omega = \frac{a (1+r^2_+/l^2)}{r^2_+ - a^2}, \quad \mu = -\frac{q r_+  }{r_+^2 - a^2}.
\end{eqnarray}
with $r_+$ defined by the equation $\Delta(r_+) = 0$. To integrate the variation of the action, we write the charges and potentials as functions of the horizon radius $r_+$, the rotational parameter $a$ and the charge $q$. One can verify that the integrability conditions hold and are satisfied by  
\begin{eqnarray}
    I = \frac{\beta}{4 l^2 \Xi} \left[ -r_+^3 + \Xi l^2 r_+ - \frac{l^2 a^2}{r_+} + \frac{l^2 q^2}{r_+} - \frac{2 l^2 q^2r_+}{r_+^2-a^2}\right]
\end{eqnarray}
which agrees with \cite{Caldarelli:1999xj} up to a sign in $a^2$ since we have kept the Euclidean signature.
\subsubsection{STU}
As our final gravitational example, we study black hole solutions in five-dimensional $\mathcal{N}=2$ gauged supergravity \cite{SABRA_1998}. Specifically, we consider the STU model \cite{Behrndt_1999}, the bosonic sector of which consists of the metric $g_{\mu\nu}$, 3 abelian gauge fields $A_{\mu}^I$ ($I=1,2,3$), and 3 real scalar fields $X^I=(S,T,U)$ satisfying the constraint $STU=1$. The Euclidean action is given by
\begin{empheq}{alignat=7}
    I[g,A,\chi]&=\frac{1}{8\pi}\int_Md^5x\sqrt{g}\Big(-\frac{1}{2}R+\frac{1}{2}G_{IJ}(X)g^{\mu\nu}\partial_\mu X^I\partial_\nu X^J+V(X)
    \cr
    &+\frac{1}{4}G_{IJ}(X)F^{I}_{\mu\nu}F^{J\mu\nu}-\frac{1}{48}C_{IJK}\epsilon^{\mu\nu\rho\sigma\lambda}F^{I}_{\mu\nu}F^{J}_{\rho\sigma}A^{K}_\lambda\Big)+B\,,
\end{empheq}
where $C_{IJK}$ is a symmetric array whose only non-vanishing component is $C_{123}=1$ and\footnote{More generally, the model is defined by a choice of constants $C_{IJK}$ and $V_I$ such that
\begin{empheq}{alignat*=7}
    \mathcal{V}(X)&\equiv\frac{1}{6}C_{IJK}X^IX^JX^K=1\,,
    &\quad
    G_{IJ}(X)&=-\left.\frac{1}{2}\partial_I\partial_J\ln{\mathcal{V}}\right|_{\mathcal{V}=1}\,,
    &\quad
    V(X)&=9V_IV_J\left(X^IX^J-\frac{1}{2}G^{IJ}\right)\,.
\end{empheq}
For the STU model, the induced metric on the surface $\mathcal{V}=1$, $g_{ij}=\partial_iX^I\partial_jX^JG_{IJ}$, is flat.
}
\begin{eqnarray}
    G_{IJ}(X)&=\frac{1}{2}\left(
    \begin{array}{*3{>{\displaystyle}c}}
        \frac{1}{S^2} & 0 & 0
        \\
        0 & \frac{1}{T^2} & 0
        \\
        0 & 0 & \frac{1}{U^2}
    \end{array}
    \right)\,,
    \hspace{9mm}
    V(X)&=\frac{2}{l^2}\left(\frac{1}{S}+\frac{1}{T}+\frac{1}{U}\right)\,.
\end{eqnarray}
The family of spherically symmetric black holes solutions we are interested in is
\begin{empheq}{alignat=7}
    ds^2&=H(r)^{-2/3}f(r)\beta^2dt^2+H(r)^{1/3}\left(\frac{dr^2}{f(r)}+r^2d\Omega^2_{3,k}\right)\,,
    \\
    A^{I}&=-i\left(\mu_I+\frac{Q_I}{r^2}H_I(r)^{-1}\right)\beta d\tau\,,
    \\
    X^I&=H(r)^{1/3}H_I(r)^{-1}\,,
\end{empheq}
with
\begin{eqnarray}
    f(r)=k-\frac{m}{r^2}+\frac{r^2}{l^2}H(r)\,, 
    \hspace{5mm}
    H_I(r)=1+\frac{q_I}{r^2}\,,
    \hspace{5mm}
    H(r)=H_1(r)H_2(r)H_3(r)\,,
\end{eqnarray}
and
\begin{eqnarray}\label{Q_I, q_I}
    Q_I^2&=kq_I^2+mq_I\,.
\end{eqnarray}
Here, $k$ parametrizes the geometry $d\Omega^2_{3,k}$ of the horizon: $k=1$ corresponds to $S^3$, $k=0$ to flat space, and $k=-1$ to the hyperbolic space $\mathds{H}^3$. Regardless, the spacetime is asymptotically $AdS_5$ with radius $l$. The solution involves $8$ integration constants, $(m,q_I)$ and $(\beta,\mu_I)$, the latter corresponding to the inverse temperature and chemical potentials, respectively.

Omitting the equations of motion for simplicity, the variation of the action reads
\begin{empheq}{alignat=7}
    \delta I&=\frac{1}{8\pi}\int_{\partial M}d^4x\sqrt{h}\,n_{\mu}\Big(-\frac{1}{2}\left(g^{\mu\rho}g^{\nu\lambda}-g^{\mu\nu}g^{\rho\lambda}\right)\nabla_{\nu}\delta g_{\rho\lambda}+G_{IJ}g^{\mu\nu}\partial_{\nu}X^I\delta X^J
    \cr
    &+G_{IJ}g^{\mu\nu}g^{\rho\sigma}F^I_{\nu\rho}\delta A^J_{\sigma}+\frac{1}{12}C_{IJK}\epsilon^{\mu\nu\rho\sigma\lambda}F^I_{\nu\rho}A^J_{\sigma}\delta A^K_{\lambda}\Big)+\delta B\,.
\end{empheq}
For the above black hole solution we find
\begin{empheq}{alignat=7}
    \delta I&=\frac{V_{3,k}}{8\pi}\left[\frac{r^4}{l^2}\delta\beta+\frac{2r^2}{3l^2}\delta(q_1\beta+q_2\beta+q_3\beta)+Q_1\delta(\mu_1\beta)+Q_2\delta(\mu_2\beta)+Q_3\delta(\mu_3\beta)\right.
    \cr
    &\left.-\frac{\beta}{2}\delta m+\left(m+\frac{q_1q_2}{3l^2}+\frac{q_1q_3}{3l^2}+\frac{q_2q_3}{3l^2}+\frac{2k}{3}q_1+\frac{2k}{3}q_2+\frac{2k}{3}q_3\right)\delta\beta\right.
    \cr
    &\left.+\left(\frac{q_2}{l^2}+\frac{q_3}{l^2}-k\right)\frac{\beta}{3}\delta q_1+\left(\frac{q_1}{l^2}+\frac{q_3}{l^2}-k\right)\frac{\beta}{3}\delta q_2+\left(\frac{q_1}{l^2}+\frac{q_2}{l^2}-k\right)\frac{\beta}{3}\delta q_3\right]+\delta B\,,
\end{empheq}
where $V_{3,k}$ denotes the volume of $d\Omega_{3,k}$. We verify once more that the divergences organize themselves into total variations, so there is no need to keep track of them. Choosing the boundary term $B$ as
\begin{empheq}{alignat=7}\label{B STU}
    B&=\frac{V_{3,k}\beta}{8\pi}\left[\frac{m}{2}+\frac{\left(q_1+q_2+q_3\right)}{3}\left(k-\frac{2r^2}{l^2}\right)-\frac{q_1q_2+q_2q_3+q_3q_1}{3l^2}-\frac{r^4}{l^2}\right]\,,
\end{empheq}
the variation becomes
\begin{empheq}{alignat=7}
    \delta I&=\left[\rho_m\delta\beta+\rho_I\delta\left(\beta\mu_I\right)\right]V_{3,k}\,,
\end{empheq}
where we recognize the mass and charge densities
\begin{empheq}{alignat=7}
    \rho_m&=\frac{1}{8\pi}\left(\frac{3m}{2}+k\left(q_1+q_2+q_3\right)\right)\,,
    \qquad
    \rho_I&=\frac{Q_I}{8\pi}\,.
\end{empheq}
It is satisfying to confirm that $\rho_m$ coincides with the ADM mass density, while $\rho_I$ is the same as the electric charge density obtained from Gauss's Law \cite{SABRA_1998}.

Regularity of the solution sets the temperature and chemical potentials to
\begin{eqnarray}
    \beta=\frac{4\pi H^{\frac{1}{2}}(r_0)}{f'(r_0)}\,,
    \hspace{8mm}
    \mu_I=-\frac{Q_I}{r_0^2H_I(r_0)}\,,
\end{eqnarray}
where $r_0$ is the location of the (outer) horizon defined by $f(r_0)=0$ (see \cite{Behrndt_1999} for a discussion on the existence of a horizon). In order to integrate the variation of the action, it is convenient to trade $m$ for $r_0$ via
\begin{eqnarray}
    m=r_0^2\left(k+\frac{r_0^2}{l^2}H(r_0)\right)\,,
\end{eqnarray}
and express $\beta$ and $\mu_I$ in terms of $r_0$ and $q_I$. One can easily check that the integrability conditions\footnote{These are equivalent to
\begin{empheq}{alignat*=7}
    \frac{\partial\rho_m}{\partial\left(\beta\mu_I\right)}&=\frac{\partial\rho_I}{\partial\beta}\,,
    &\qquad
    \frac{\partial\rho_I}{\partial\left(\beta\mu_J\right)}&=\frac{\partial\rho_J}{\partial\left(\beta\mu_I\right)}\,.
\end{empheq}
}
\begin{empheq}{alignat=7}
    \frac{\partial}{\partial q_I}\left(\rho_m\frac{\partial\beta}{\partial r_0}+\rho_J\frac{\partial\left(\beta\mu_J\right)}{\partial r_0}\right)&=\frac{\partial}{\partial r_0}\left(\rho_m\frac{\partial\beta}{\partial q_I}+\rho_J\frac{\partial\left(\beta\mu_J\right)}{\partial q_I}\right)\,,
    \\
    \frac{\partial}{\partial q_I}\left(\rho_m\frac{\partial\beta}{\partial q_J}+\rho_K\frac{\partial\left(\beta\mu_K\right)}{\partial q_J}\right)&=\frac{\partial}{\partial q_J}\left(\rho_m\frac{\partial\beta}{\partial q_I}+\rho_K\frac{\partial\left(\beta\mu_K\right)}{\partial q_I}\right)\,,
\end{empheq}
are satisfied and solved by
\begin{eqnarray}
    I[\beta,\mu]=\frac{\beta r_0^2}{16\pi}\left(k-\frac{r_0^2}{l^2}H(r_0)\right)V_{3,k}\,.
\end{eqnarray}
The entropy then becomes
\begin{empheq}{alignat=7}
    S[M,Q]&=\beta\left(\rho_m+\mu_I\rho_I\right)V_{3,k}-I[\beta,\mu]
    \cr
    &=\frac{H^{\frac{1}{2}}(r_0)r_0^3V_{3,k}}{4}\,.
\end{empheq}
This is $\frac{A}{4}$.
\subsection{Holographic Wilson loops}
\label{WilsonSec}
The variational methods developed in this paper are not limited to gravitational actions. With a few modifications, they can be applied in a variety of problems. Here we discuss two examples borrowed from the holographic Wilson loop literature: the quark-antiquark potential and the circular Wilson loop. 

According to the AdS/CFT correspondence, a Wilson loop operator is dual to a macroscopic string whose worldsheet pinches the boundary of AdS along the loop \cite{Rey:1998ik,Maldacena:1998im}. At strong coupling, the expectation value of the Wilson loop is given by
\begin{empheq}{alignat=7}\label{WL dictionary}
    \langle W\rangle&\sim e^{-S}\,,
\end{empheq}
where $S$ is the on-shell action of the string. Our methods can be applied to compute such action.

\subsubsection{Quark-antiquark potential}
It is a standard result in Quantum Field Theory that the expectation value of a rectangular Wilson loop can be interpreted in terms of the effective potential $V(L)$ between two probe particles. Indeed,
\begin{empheq}{alignat=7}
    \langle W_{\textrm{rect}}\rangle&\underset{T\to\infty}{\sim}e^{-TV(L)}\,,
\end{empheq}
where $T$ is the length of the rectangle along the (Euclidean) time direction and $L$ its width. The duality \eqref{WL dictionary} then states that
\begin{empheq}{alignat=7}\label{Vqq}
    V(L)&=\lim_{T\to\infty}\frac{S(L)}{T}\,.
\end{empheq}
Our method will deliver the derivative of the potential with respect to $L$.
\begin{figure}[h]
    \centering
    \includegraphics[width=0.75\textwidth]{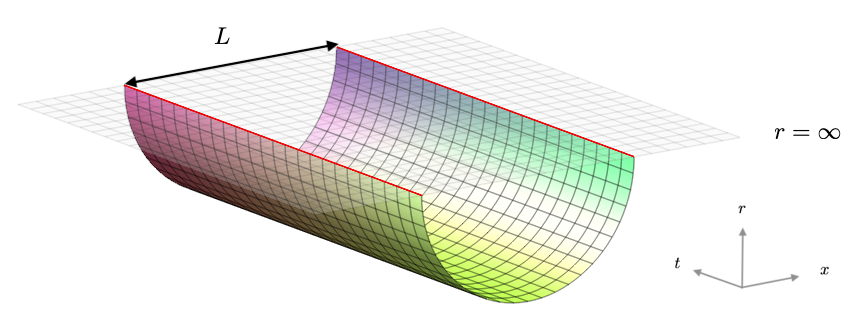}
    \caption{Caption}
    \label{fig: rectangular}
\end{figure}

Although we could consider more general supergravity backgrounds,  it will be sufficient for our purposes to consider a string propagating on the asymptotically $AdS_3$ space
\begin{empheq}{alignat=7}
    ds^2&=f(r)\left(dt^2+dx^2\right)+\frac{dr^2}{f(r)}\,,
    &\qquad
    f(r)&=\frac{r^2}{R^2}+\cdots\,.
\end{empheq}
It is important to emphazise that the function $f(r)$ is fixed and will not be varied in what follows. Using $t$ and $r$ as worldsheet coordinates, the embedding of the string can be described by a single function $x(r)$.
This only covers half of the worlsheet (see figure \ref{fig: rectangular}), up to the turning point $r=r_*$. The other half is given by the mirror image $x(r)\to-x(r)$. The Nambu-Goto action for the string then becomes
\begin{empheq}{alignat=7}\label{NG gauge fixed}
    S[x]&=\frac{T}{\pi\alpha'}\int_{r_*}^{\infty}dr\,\sqrt{1+f(r)^2x'(r)^2}+B\,.
\end{empheq}
We have included a factor of $2$ to account for both halves of the worldsheet. The boundary term $B$ must be chosen such that the action is finite and satisfies the variational problem with fixed separation $L$.

Having identified the relevant action, the next step is to characterize the space of classical solutions. Since the Lagrangian does not depend on $x(r)$, the corresponding momentum $p$ is conserved, allowing us to write the first order equation
\begin{empheq}{alignat=7}
    x'(r)&=\frac{p}{f(r)\sqrt{f(r)^2-p^2}}\,.
\end{empheq}
We find that solutions are parametrized by the pair of integrtaion constants $(L,p)$ and have the asymptotic expansion
\begin{empheq}{alignat=7}
     x(r)&\underset{r\to\infty}{\longrightarrow}\frac{L}{2}-\frac{pR^4}{3r^3}+\cdots\,.
\end{empheq}
This reveals the role of $L$ as the leading term (source/chemical potential) and $p$ as sub-leading (vev/charge). Notice that we can (implicitly) eliminate the turning point $r_*$ in favor of $L$ and $p$ through the boundary condition
\begin{empheq}{alignat=7}\label{eq: turning point}
    \int_{r_*}^{\infty}drx'(r)&=\frac{L}{2}\,.
\end{empheq}
Now, the on-shell variation of the action yields
\begin{empheq}{alignat=7}\label{dSqq}
    \delta S&=\frac{T}{2\pi\alpha'}\,p\delta L+\delta B\,.
\end{empheq}
As it turns out, this is finite and requires no renormalization. We also confirm that $L$ and $p$ are conjugate variables and that the action is already suited for the Dirichlet problem. Thus, $B=0$. Finally, the regularity condition relating the two integration constants is
\begin{empheq}{alignat=7}
    \frac{1}{x'(r_*)}&=0
    &\qquad\Rightarrow\qquad
    p&=f(r_*)\,,
\end{empheq}
meaning that the string is horizontal at the turning point. 

Putting \eqref{Vqq} and \eqref{dSqq} together, we find that the derivative of the quark-antiquark potential is
\begin{empheq}{alignat=7}
    \frac{dV}{dL}&=\frac{f(r_*)}{2\pi\alpha'}\,.
\end{empheq}
This result was derived in \cite{Nunez:2009da} using background subtraction. It is now a technical matter to find $r_*$ as a function of $L$. For the exact $AdS$ background relation \eqref{eq: turning point} becomes
\begin{empheq}{alignat=7}
    L&=\frac{R^2}{r_*}\Gamma\left(\frac{3}{4}\right)^2\sqrt{\frac{2}{\pi}}\,.
\end{empheq}
Thus,
\begin{empheq}{alignat=7}
    \frac{dV}{dL}&=\frac{\Gamma\left(\frac{3}{4}\right)^4}{\pi^2\alpha'}\frac{R^2}{L^2}
    &\qquad\Rightarrow\qquad
    V(L)&=-\frac{\Gamma\left(\frac{3}{4}\right)^4}{\pi^2}\frac{\sqrt{\lambda}}{L}\,,
\end{empheq}
where we have used the relation $R^2=\alpha'\sqrt{\lambda}$ between the $AdS$ radius and the 't Hooft coupling in the gauge theory.
\subsubsection{Circular Wilson loop}
The other seminal example is the $\frac{1}{2}$-BPS circular Wilson loop. This case is interesting because it illustrates the limitations of our method; due to the residual conformal invariance preserved by the circle, the vev of the Wilson loop is constant, independent of the radius, and cannot be determined by looking uniquely at the variation of the action. This ambiguity, however, is present in any renomalization scheme.

Given the symmetries of the problem, we write the $AdS_3$ background as
\begin{empheq}{alignat*=7}
    ds^2&=\frac{R^2}{z^2}\left(dz^2+d\rho^2+\rho^2d\phi^2\right)\,.
\end{empheq}
Choosing $z$ and $\phi$ as worldsheet coordinates, the string embedding is characterized by a function $\rho(z)$ such that $\rho(z\to0)\to\rho_0$, where $\rho_0$ is the radius of the loop. The dynamics is governed by the Nambu-Goto action
\begin{empheq}{alignat*=7}
    S[\rho]&=\frac{R^2}{\alpha'}\int_{\epsilon}^{z_*}dz\,\frac{\rho}{z^2}\sqrt{1+\rho'^2}+B\,,
\end{empheq}
where $\epsilon\ll1$ is a small regulator and $z_*$ is the turning point determined by the regularity condition
\begin{empheq}{alignat*=7}
    \frac{1}{\rho'(z_*)}&=0\,.
\end{empheq}
The boundary term $B$ must render a finite variation $\delta S\propto\delta\rho_0$.

It is hard to find the exact general solution in this case, but for $z\to0$ the asymptotic expansion is
\begin{empheq}{alignat*=7}
    \rho(z)&=\rho_0+\rho_1z+\rho_2z^2+\cdots\,.
\end{empheq}
The equation of motion leaves $\rho_0$ and $\rho_3$ free and fixes
\begin{empheq}{alignat*=7}
    \rho_1&=0\,,
    &\qquad
    \rho_2&=-\frac{1}{2\rho_0}\,,
    &\qquad
    \cdots\,.
\end{empheq}
For the variation of the action we find
\begin{empheq}{alignat*=7}
    \delta S&=\frac{R^2}{\alpha'}\left(-\frac{\delta\rho_0}{z}+3\rho_0\rho_3\delta\rho_0\right)+\delta B\,.
\end{empheq}
Here there is a divergent term, but, as expected, it is a total variation. Moreover, we verify once more that the divergence satisfies
\begin{empheq}{alignat=7}
    \delta\left(\frac{\rho}{z^2}\sqrt{1+\rho'^2}\right)&\underset{z\to0}{\longrightarrow}\frac{d}{dz}\left(-\frac{\delta\rho_0}{z}\right)\,.
\end{empheq}
Choosing
\begin{empheq}{alignat=7}\label{B circular}
    B&=\frac{R^2}{\alpha'}\frac{\rho_0}{\epsilon}\,,
\end{empheq}
the variation becomes
\begin{empheq}{alignat=7}\label{dS circular}
    \delta S&=\frac{3R^2}{\alpha'}\rho_0\rho_3\delta\rho_0\,.
\end{empheq}
Notice that the counterterm \eqref{B circular} can be written as\footnote{To compute $\Pi_{z}=\frac{\partial L}{\partial z'}$ one must undo the gauge-fixing and introduce a new worldsheet coordinate $\sigma$ such that $z=z(\sigma)$ and $\rho=\rho(\sigma)$. After computing the momentum, one can gauge-fix again and set $z'(\sigma)=-1$.}
\begin{empheq}{alignat=7}
    B&=z\Pi_z\Big|_{z=\epsilon}\,,
    &\qquad
    \Pi_{z}&=\frac{R^2}{\alpha'}\frac{\rho}{z^2\sqrt{1+\rho'^2}}\,.
\end{empheq}
This coincides with the standard renormalization prescription whereby divergences are removed by a Legendre transform. Finally, the regular solution to the equation of motion is known exactly and reads
\begin{empheq}{alignat*=7}
    \rho(z)&=\sqrt{\rho_0^2-z^2}&&=\rho_0\left(1-\frac{1}{2}\frac{z^2}{\rho_0^2}-\frac{1}{8}\frac{z^4}{\rho_0^4}+\cdots\right)\,.
\end{empheq}
We see that regularity imposes $\rho_3=0$, so the variation \eqref{dS circular} vanishes. Of course, the fact that the action does not depend on $\rho_0$ is a consequence of conformal invariance. It is interesting to notice that since the counterterm $B$ does not have a finite contribution, the actual value of the renormalized action comes entirely from the bulk integral. Indeed,
\begin{empheq}{alignat=7}
    S&=\frac{R^2}{\alpha'}\int_{\epsilon}^{z_*}dz\,\frac{\rho_0}{z^2}+\frac{R^2}{\alpha'}\frac{\rho_0}{\epsilon}
    \cr
    &=-\frac{R^2}{\alpha'}\,,
\end{empheq}
where we have used that $z_*=\rho_0$.
\section{Conclusions}

To conclude, in this paper we have presented a method for computing on-shell actions. The method is general, although we have focus on black holes and holographic Wilson loops.  

The main advantage of this method is that divergencies can be treated in a universal way. In particular, we have argued that the correct finite part of the action can be extracted without having to compute the divergent part. This follows from the existence of extra structure in the action variation, namely the distinction between exact and non-exact 1-forms on field space. 

We have applied the method to several examples with varying degrees of complexity. The main focus has been to stress that, within the variational approach,  IR divergencies do not play a role when extracting the finite part of the action.


\appendix
\section{Acknowledgments}

MB thanks R. Emparan for an instructive conversation on holographic renormalization, and L. Garay for hospitality and stimulating discussions at U. Complutense (Madrid), where this work was initiated. Useful conversations with G. Barnich and C. Corral are also acknowledged.  DA acknowledges financial support from the Agencia Nacional de Investigación y Desarrollo (ANID)-Scholarship Program through the Doctorado Nacional Grant No. 2023-21231555.

\bibliographystyle{unsrt}
\bibliography{references}

@Article{Witten:1998qj,
  author        = {Witten, Edward},
  journal       = {Adv. Theor. Math. Phys.},
  title         = {{Anti de Sitter space and holography}},
  year          = {1998},
  pages         = {253--291},
  volume        = {2},
  archiveprefix = {arXiv},
  doi           = {10.4310/ATMP.1998.v2.n2.a2},
  eprint        = {hep-th/9802150},
  reportnumber  = {IASSNS-HEP-98-15},
}

@Article{Gubser:1998bc,
  author        = {Gubser, S. S. and Klebanov, Igor R. and Polyakov, Alexander M.},
  journal       = {Phys. Lett. B},
  title         = {{Gauge theory correlators from noncritical string theory}},
  year          = {1998},
  pages         = {105--114},
  volume        = {428},
  archiveprefix = {arXiv},
  doi           = {10.1016/S0370-2693(98)00377-3},
  eprint        = {hep-th/9802109},
  reportnumber  = {PUPT-1767},
}

@Article{Maldacena:1997re,
  author        = {Maldacena, Juan Martin},
  journal       = {Adv. Theor. Math. Phys.},
  title         = {{The Large $N$ limit of superconformal field theories and supergravity}},
  year          = {1998},
  pages         = {231--252},
  volume        = {2},
  archiveprefix = {arXiv},
  doi           = {10.4310/ATMP.1998.v2.n2.a1},
  eprint        = {hep-th/9711200},
  reportnumber  = {HUTP-97-A097, HUTP-98-A097},
}

@Article{Balasubramanian:1999re,
  author        = {Balasubramanian, Vijay and Kraus, Per},
  journal       = {Commun. Math. Phys.},
  title         = {{A Stress tensor for Anti-de Sitter gravity}},
  year          = {1999},
  pages         = {413--428},
  volume        = {208},
  archiveprefix = {arXiv},
  doi           = {10.1007/s002200050764},
  eprint        = {hep-th/9902121},
  reportnumber  = {HUTP-99-A002, EFI-99-6, NSF-ITP-98-132},
}

@article{Astefanesei:2004ji,
    author = "Astefanesei, Dumitru and Mann, Robert B. and Radu, Eugen",
    title = "{Breakdown of the entropy/area relationship for NUT-charged spacetimes}",
    eprint = "hep-th/0406050",
    archivePrefix = "arXiv",
    doi = "10.1016/j.physletb.2005.05.057",
    journal = "Phys. Lett. B",
    volume = "620",
    pages = "1--8",
    year = "2005"
}

@Article{Banados:1993qp,
  author        = {Bañados, Maximo and Teitelboim, Claudio and Zanelli, Jorge},
  journal       = {Phys. Rev. Lett.},
  title         = {{Black hole entropy and the dimensional continuation of the Gauss-Bonnet theorem}},
  year          = {1994},
  pages         = {957--960},
  volume        = {72},
  archiveprefix = {arXiv},
  doi           = {10.1103/PhysRevLett.72.957},
  eprint        = {gr-qc/9309026},
  reportnumber  = {IASSNS-HEP-93-53},
}

@Article{Jacobson:1993xs,
  author        = {Jacobson, Ted and Myers, Robert C.},
  journal       = {Phys. Rev. Lett.},
  title         = {{Black hole entropy and higher curvature interactions}},
  year          = {1993},
  pages         = {3684--3687},
  volume        = {70},
  archiveprefix = {arXiv},
  doi           = {10.1103/PhysRevLett.70.3684},
  eprint        = {hep-th/9305016},
  reportnumber  = {NSF-ITP-93-41, MCGILL-93-04, UMDGR-93-179},
}

@article{Brown:1992bq,
    author = "Brown, J. David and York, Jr., James W.",
    title = "{The Microcanonical functional integral. 1. The Gravitational field}",
    eprint = "gr-qc/9209014",
    archivePrefix = "arXiv",
    reportNumber = "IFP-441-UNC, TAR-028-UNC",
    doi = "10.1103/PhysRevD.47.1420",
    journal = "Phys. Rev. D",
    volume = "47",
    pages = "1420--1431",
    year = "1993"
}

@article{Brown:1990fk,
    author = "Brown, J. David and Martinez, Erik A. and York, Jr., James W.",
    title = "{Complex Kerr-Newman geometry and black hole thermodynamics}",
    reportNumber = "IFP-1000, IFP-1001",
    doi = "10.1103/PhysRevLett.66.2281",
    journal = "Phys. Rev. Lett.",
    volume = "66",
    pages = "2281--2284",
    year = "1991"
}

@InBook{Hawking:1980gf,
  author    = {Hawking, S. W.},
  pages     = {746--789},
  publisher = {Cambride University Press},
  title     = {{The Path-Integral Approach to Quantum Gravity}},
  year      = {1979},
  booktitle = {{General Relativity}: {An Einstein Centenary Survey}},
}

@article{Gutperle:2011kf,
    author = "Gutperle, Michael and Kraus, Per",
    title = "{Higher Spin Black Holes}",
    eprint = "1103.4304",
    archivePrefix = "arXiv",
    primaryClass = "hep-th",
    doi = "10.1007/JHEP05(2011)022",
    journal = "JHEP",
    volume = "05",
    pages = "022",
    year = "2011"
}

@article{Banados:2012ue,
    author = "Banados, Maximo and Canto, Rodrigo and Theisen, Stefan",
    title = "{The Action for higher spin black holes in three dimensions}",
    eprint = "1204.5105",
    archivePrefix = "arXiv",
    primaryClass = "hep-th",
    doi = "10.1007/JHEP07(2012)147",
    journal = "JHEP",
    volume = "07",
    pages = "147",
    year = "2012"
}

@Article{Banados:2004zt,
  author        = {Banados, M. and Schwimmer, A. and Theisen, S.},
  journal       = {JHEP},
  title         = {{Chern-Simons gravity and holographic anomalies}},
  year          = {2004},
  pages         = {039},
  volume        = {05},
  archiveprefix = {arXiv},
  doi           = {10.1088/1126-6708/2004/05/039},
  eprint        = {hep-th/0404245},
  reportnumber  = {AEI-2004-034},
}

@Article{Brown:1989fa,
  author       = {Brown, J. David and Comer, G. L. and Martinez, E. A. and Melmed, J. and Whiting, Bernard F. and York, Jr., James W.},
  journal      = {Class. Quant. Grav.},
  title        = {{Thermodynamic Ensembles and Gravitation}},
  year         = {1990},
  pages        = {1433--1444},
  volume       = {7},
  doi          = {10.1088/0264-9381/7/8/020},
  reportnumber = {IFP-357-UNC},
}

@Article{Brown:1992br,
  author        = {Brown, J. David and York, Jr., James W.},
  journal       = {Phys. Rev. D},
  title         = {{Quasilocal energy and conserved charges derived from the gravitational action}},
  year          = {1993},
  pages         = {1407--1419},
  volume        = {47},
  archiveprefix = {arXiv},
  doi           = {10.1103/PhysRevD.47.1407},
  eprint        = {gr-qc/9209012},
  reportnumber  = {IFP-423-UNC, TAR-009-UNC},
}

@Article{Banados:1992wn,
  author        = {Bañados, Maximo and Teitelboim, Claudio and Zanelli, Jorge},
  journal       = {Phys. Rev. Lett.},
  title         = {{The Black hole in three-dimensional space-time}},
  year          = {1992},
  pages         = {1849--1851},
  volume        = {69},
  archiveprefix = {arXiv},
  doi           = {10.1103/PhysRevLett.69.1849},
  eprint        = {hep-th/9204099},
  reportnumber  = {PRINT-92-0151 (CHILE), IASSNS-HEP-92-29},
}

@article{Bahamonde:2025qtc,
    author = "Bahamonde, Sebastian and Ba{\~n}ados, M{\'a}ximo",
    title = "{An exact five dimensional Weyl-geometry Gauss-Bonnet black hole}",
    eprint = "2504.02230",
    archivePrefix = "arXiv",
    primaryClass = "gr-qc",
    doi = "10.1016/j.physletb.2025.139869",
    journal = "Phys. Lett. B",
    volume = "869",
    pages = "139869",
    year = "2025"
}

@Article{Gibbons:2004ai,
  author        = {Gibbons, G. W. and Perry, M. J. and Pope, C. N.},
  journal       = {Class. Quant. Grav.},
  title         = {{The First law of thermodynamics for Kerr-anti-de Sitter black holes}},
  year          = {2005},
  pages         = {1503--1526},
  volume        = {22},
  archiveprefix = {arXiv},
  doi           = {10.1088/0264-9381/22/9/002},
  eprint        = {hep-th/0408217},
  reportnumber  = {DAMTP-2004-87, MIFP-04-17},
}

@Article{Barnich:2004uw,
  author        = {Barnich, G. and Compere, G.},
  journal       = {Phys. Rev. D},
  title         = {{Generalized Smarr relation for Kerr AdS black holes from improved surface integrals}},
  year          = {2005},
  note          = {[Erratum: Phys.Rev.D 73, 029904 (2006)]},
  pages         = {044016},
  volume        = {71},
  archiveprefix = {arXiv},
  doi           = {10.1103/PhysRevD.73.029904},
  eprint        = {gr-qc/0412029},
  reportnumber  = {ULB-TH-04-31},
}

@Article{Chamblin:1998pz,
  author        = {Chamblin, Andrew and Emparan, Roberto and Johnson, Clifford V. and Myers, Robert C.},
  journal       = {Phys. Rev. D},
  title         = {{Large N phases, gravitational instantons and the nuts and bolts of AdS holography}},
  year          = {1999},
  pages         = {064010},
  volume        = {59},
  archiveprefix = {arXiv},
  doi           = {10.1103/PhysRevD.59.064010},
  eprint        = {hep-th/9808177},
}

@Article{Andrade:2006pg,
  author        = {Andrade, Tomas and Banados, Maximo and Rojas, Francisco},
  journal       = {Phys. Rev. D},
  title         = {{Variational Methods in AdS/CFT}},
  year          = {2007},
  pages         = {065013},
  volume        = {75},
  archiveprefix = {arXiv},
  doi           = {10.1103/PhysRevD.75.065013},
  eprint        = {hep-th/0612150},
}

@Article{Crnkovic:1986ex,
  author       = {Crnkovic, Cedomir and Witten, Edward},
  journal      = {300 Years of Gravitation (edited by S. Hawking and W. Israel), Cambridge University Press},
  title        = {{Covariant description of canonical formalism in geometrical theories}},
  year         = {1987},
  month        = {9},
  reportnumber = {Print-86-1309 (PRINCETON)},
}

@Article{Witten:1998zw,
  author        = {Witten, Edward},
  journal       = {Adv. Theor. Math. Phys.},
  title         = {{Anti-de Sitter space, thermal phase transition, and confinement in gauge theories}},
  year          = {1998},
  pages         = {505--532},
  volume        = {2},
  archiveprefix = {arXiv},
  doi           = {10.4310/ATMP.1998.v2.n3.a3},
  editor        = {Bergstrom, L. and Lindstrom, U.},
  eprint        = {hep-th/9803131},
  reportnumber  = {IASSNS-HEP-98-21},
}

@ARTICLE{Gibbons:1976ue,
  author = {Gibbons, G. W. and Hawking, S. W.},
  title = {{Action Integrals and Partition Functions in Quantum Gravity}},
  journal = {Phys. Rev.},
  year = {1977},
  volume = {D15},
  pages = {2752-2756},
  doi = {10.1103/PhysRevD.15.2752},
  slaccitation = {%%CITATION = PHRVA,D15,2752;%%}
}

@Article{deHaro:2000vlm,
  author        = {de Haro, Sebastian and Solodukhin, Sergey N. and Skenderis, Kostas},
  journal       = {Commun. Math. Phys.},
  title         = {{Holographic reconstruction of space-time and renormalization in the AdS / CFT correspondence}},
  year          = {2001},
  pages         = {595--622},
  volume        = {217},
  archiveprefix = {arXiv},
  doi           = {10.1007/s002200100381},
  eprint        = {hep-th/0002230},
  reportnumber  = {SPIN-2000-05, ITP-UU-00-03, PUTP-1921},
}

@Article{Papadimitriou:2005ii,
  author        = {Papadimitriou, Ioannis and Skenderis, Kostas},
  journal       = {JHEP},
  title         = {{Thermodynamics of asymptotically locally AdS spacetimes}},
  year          = {2005},
  pages         = {004},
  volume        = {08},
  archiveprefix = {arXiv},
  doi           = {10.1088/1126-6708/2005/08/004},
  eprint        = {hep-th/0505190},
  reportnumber  = {ITFA-2005-18},
}

@Article{Wald:1993nt,
  author        = {Wald, Robert M.},
  journal       = {Phys. Rev. D},
  title         = {{Black hole entropy is the Noether charge}},
  year          = {1993},
  number        = {8},
  pages         = {R3427--R3431},
  volume        = {48},
  archiveprefix = {arXiv},
  doi           = {10.1103/PhysRevD.48.R3427},
  eprint        = {gr-qc/9307038},
  reportnumber  = {EFI-93-42},
}

@ARTICLE{ReggeTeitelboim,
  author = {Regge, Tullio and Teitelboim, Claudio},
  title = {{Role of Surface Integrals in the Hamiltonian Formulation of General
	Relativity}},
  journal = {Ann. Phys.},
  year = {1974},
  volume = {88},
  pages = {286},
  doi = {10.1016/0003-4916(74)90404-7},
  slaccitation = {%%CITATION = APNYA,88,286;%%}
}

@Article{Barnich:2001jy,
  author        = {Barnich, Glenn and Brandt, Friedemann},
  journal       = {Nucl.Phys.},
  title         = {{Covariant theory of asymptotic symmetries, conservation laws and central charges}},
  year          = {2002},
  pages         = {3-82},
  volume        = {B633},
  archiveprefix = {arXiv},
  doi           = {10.1016/S0550-3213(02)00251-1},
  eprint        = {hep-th/0111246},
  primaryclass  = {hep-th},
}

@Article{Iyer:1994ys,
  author        = {Iyer, Vivek and Wald, Robert M.},
  journal       = {Phys. Rev. D},
  title         = {{Some properties of Noether charge and a proposal for dynamical black hole entropy}},
  year          = {1994},
  pages         = {846--864},
  volume        = {50},
  archiveprefix = {arXiv},
  doi           = {10.1103/PhysRevD.50.846},
  eprint        = {gr-qc/9403028},
}

@Article{Maldacena:2016upp,
  author        = {Maldacena, Juan and Stanford, Douglas and Yang, Zhenbin},
  journal       = {PTEP},
  title         = {{Conformal symmetry and its breaking in two dimensional Nearly Anti-de-Sitter space}},
  year          = {2016},
  number        = {12},
  pages         = {12C104},
  volume        = {2016},
  archiveprefix = {arXiv},
  doi           = {10.1093/ptep/ptw124},
  eprint        = {1606.01857},
  primaryclass  = {hep-th},
}

@article{SABRA_1998,
   title={GENERAL BPS BLACK HOLES IN FIVE DIMENSIONS},
   volume={13},
   ISSN={1793-6632},
   url={http://dx.doi.org/10.1142/S0217732398000309},
   DOI={10.1142/s0217732398000309},
   number={03},
   journal={Modern Physics Letters A},
   publisher={World Scientific Pub Co Pte Lt},
   author={Sabra, W. A.},
   year={1998},
   month=jan, pages={239–252} }

@article{Behrndt_1999,
   title={Non-extreme black holes of five-dimensional N = 2 AdS supergravity},
   volume={553},
   ISSN={0550-3213},
   url={http://dx.doi.org/10.1016/S0550-3213(99)00243-6},
   DOI={10.1016/s0550-3213(99)00243-6},
   number={1–2},
   journal={Nuclear Physics B},
   publisher={Elsevier BV},
   author={Behrndt, K. and Cvetič, M. and Sabra, W.A.},
   year={1999},
   month=jul, pages={317–332} }

@article{Mann:1999pc,
    author = "Mann, Robert B.",
    title = "{Misner string entropy}",
    eprint = "hep-th/9903229",
    archivePrefix = "arXiv",
    reportNumber = "WATPHYS-TH99-01",
    doi = "10.1103/PhysRevD.60.104047",
    journal = "Phys. Rev. D",
    volume = "60",
    pages = "104047",
    year = "1999"
}

@article{Emparan:1999pm,
    author = "Emparan, Roberto and Johnson, Clifford V. and Myers, Robert C.",
    title = "{Surface terms as counterterms in the AdS / CFT correspondence}",
    eprint = "hep-th/9903238",
    archivePrefix = "arXiv",
    reportNumber = "DTP-99-21, UK-99-04, MCGILL-99-12, EHU-FT-9906",
    doi = "10.1103/PhysRevD.60.104001",
    journal = "Phys. Rev. D",
    volume = "60",
    pages = "104001",
    year = "1999"
}

@article{Hawking:1998jf,
    author = "Hawking, S. W. and Hunter, C. J.",
    title = "{Gravitational entropy and global structure}",
    eprint = "hep-th/9808085",
    archivePrefix = "arXiv",
    reportNumber = "DAMTP-98-104",
    doi = "10.1103/PhysRevD.59.044025",
    journal = "Phys. Rev. D",
    volume = "59",
    pages = "044025",
    year = "1999"
}

@article{PhysRevD.59.044033,
  title = {NUT charge, anti--de Sitter space, and entropy},
  author = {Hawking, S. W. and Hunter, C. J. and Page, Don N.},
  journal = {Phys. Rev. D},
  volume = {59},
  issue = {4},
  pages = {044033},
  numpages = {6},
  year = {1999},
  month = {Jan},
  publisher = {American Physical Society},
  doi = {10.1103/PhysRevD.59.044033},
  url = {https://link.aps.org/doi/10.1103/PhysRevD.59.044033}
}

@article{Nunez:2009da,
    author = "Nunez, Carlos and Piai, Maurizio and Rago, Antonio",
    title = "{Wilson Loops in string duals of Walking and Flavored Systems}",
    eprint = "0909.0748",
    archivePrefix = "arXiv",
    primaryClass = "hep-th",
    doi = "10.1103/PhysRevD.81.086001",
    journal = "Phys. Rev. D",
    volume = "81",
    pages = "086001",
    year = "2010"
}

@article{Caldarelli:1999xj,
    author = "Caldarelli, Marco M. and Cognola, Guido and Klemm, Dietmar",
    title = "{Thermodynamics of Kerr-Newman-AdS black holes and conformal field theories}",
    eprint = "hep-th/9908022",
    archivePrefix = "arXiv",
    reportNumber = "UTF-434",
    doi = "10.1088/0264-9381/17/2/310",
    journal = "Class. Quant. Grav.",
    volume = "17",
    pages = "399--420",
    year = "2000"
}

@article{Rey:1998ik,
    author = "Rey, Soo-Jong and Yee, Jung-Tay",
    title = "{Macroscopic strings as heavy quarks in large N gauge theory and anti-de Sitter supergravity}",
    eprint = "hep-th/9803001",
    archivePrefix = "arXiv",
    reportNumber = "SNUTP-98-016",
    doi = "10.1007/s100520100799",
    journal = "Eur. Phys. J. C",
    volume = "22",
    pages = "379--394",
    year = "2001"
}

@article{Maldacena:1998im,
    author = "Maldacena, Juan Martin",
    title = "{Wilson loops in large N field theories}",
    eprint = "hep-th/9803002",
    archivePrefix = "arXiv",
    reportNumber = "HUTP-98-A014",
    doi = "10.1103/PhysRevLett.80.4859",
    journal = "Phys. Rev. Lett.",
    volume = "80",
    pages = "4859--4862",
    year = "1998"
}
\end{document}